\documentclass[prb,twocolumn,a4paper,showpacs,aps,superscriptaddress]{revtex4}
\pdfoutput=1
\usepackage{amsmath}
\usepackage{amsfonts}
\usepackage{dsfont}
\usepackage{graphicx}
\usepackage{psfrag}
\begin{document}
\title{Transport via coupled states in a C$_{60}$ peapod quantum dot}
\author{Anders Eliasen}
\author{Jens Paaske}
\email[author to whom correspondence should be addressed:]{paaske@fys.ku.dk}
\author{Karsten Flensberg}
\affiliation{Niels Bohr Institute \& Nano-Science Center,
University of Copenhagen, Universitetsparken 5, 2100~Copenhagen \O ,
Denmark}
\author{Sebastian Smerat}
\affiliation{Ludwig-Maximilians-Universit\"at M\"unchen, Physics
Department, Arnold Sommerfeld Center for Theoretical Physics, and
Center for NanoScience, Ludwig-Maximilians-Universit\"at M\"unchen,
D-80333 M\"unchen, Germany}
\affiliation{Institut f\"ur Theoretische Physik A, RWTH Aachen,
52056 Aachen, Germany}
\affiliation{JARA- Fundamentals of Future Information Technology}
\author{Martin Leijnse}
\affiliation{Niels Bohr Institute \& Nano-Science Center,
University of Copenhagen, Universitetsparken 5, 2100~Copenhagen \O ,
Denmark}
\affiliation{Institut f\"ur Theoretische Physik A, RWTH Aachen,
52056 Aachen, Germany}
\affiliation{JARA- Fundamentals of Future Information Technology}
\author{Maarten R. Wegewijs}
\affiliation{Institut f\"ur Festk{\"o}rperforschung,
Forschungszentrum J{\"u}lich, 524 25 J{\"u}lich,  Germany}
\affiliation{Institut f\"ur Theoretische Physik A, RWTH Aachen,
52056 Aachen, Germany}
\affiliation{JARA- Fundamentals of Future Information Technology}
\author{Henrik I. J{\o}rgensen}
\affiliation{Niels Bohr Institute \& Nano-Science Center,
University of Copenhagen, Universitetsparken 5, 2100~Copenhagen \O ,
Denmark}
\author{Marc Monthioux}
\affiliation{CEMES, Centre d'Elaboration des Mat\'{e}riaux et d'Etudes
Structurales (CEMES), UPR A-8011 CNRS, B.P. 94347, 29 Rue Jeanne Marvig,
F-31055 Toulouse Cedex 4, France}
\author{Jesper Nyg{\aa}rd}
\affiliation{Niels Bohr Institute \& Nano-Science Center,
University of Copenhagen, Universitetsparken 5, 2100~Copenhagen \O ,
Denmark}

\date{\today}

\begin{abstract}
We have measured systematic repetitions of avoided crossings in low
temperature three-terminal transport through a carbon nanotube with
encapsulated C$_{60}$ molecules. We show that this is a general
effect of the hybridization of a host quantum dot with an impurity.
The well-defined nanotube allows identification of the properties of
the impurity, which we suggest to be a chain of C$_{60}$ molecules
inside the nanotube. This electronic coupling between the two
subsystems opens the interesting and potentially useful possibility
of contacting the encapsulated molecules via the tube.

\end{abstract}


\maketitle


Peapod systems represent a next step in complexity of carbon-based
electronics, departing from the well-characterized single-walled
nanotube system. Since the advent~\cite{Smith98} of these
single-walled carbon nanotubes (CNTs) filled with C$_{60}$ molecules
(or other fullerenes), there has been an ongoing experimental effort
to clarify the modification of the electronic properties of the CNT.
In particular, the hybridization of the C$_{60}$ molecules with the
CNT electronic states is of importance for addressing the molecular
scale "peas" via the CNT, for instance using spin-exchange
processes. More generally, the interaction of quantum dot systems of
different nature and the associated transport signatures are of
broad interest. Band structure
calculations~\cite{Okada01,Kane02,Lu03,Yoon03,Dubay04,Kondo05} have
suggested that the hybridization between the CNT and the
encapsulated C$_{60}$ molecules could lead to an extra band crossing
the Fermi-level in a metallic peapod, depending on tube-chirality,
but the experimental evidence for such mixing between the two
subsystems remains ambiguous.

Since the first transmission electron microscopy images of the
encapsulated molecules~\cite{Smith98}, scanning tunneling microscopy
(STM) has been used to probe the electronic states of a single
peapod, showing that they were indeed different from those of an
empty CNT~\cite{Hornbaker02}. These STM-data of Hornbaker {\it et
al.}~\cite{Hornbaker02} were rationalized in terms of a
semi-empirical model invoking a coupling between the CNT
$\pi$-orbitals and the $t_{1u}$ states of a C$_{60}$ of the order of
$1.25$~eV, indicative of substantial hybridization of the two
subsystems. DFT-results of Lu {\it et al.}~\cite{Lu03} predicted one
order of magnitude smaller hybridization. Subsequent photoemission
studies~\cite{Shiozawa06} even showed no evidence for hybridization
between C$_{60}$ molecules and tube.

Also low-temperature transport measurements of peapods prepared as
three-terminal quantum dots have been
performed~\cite{Utko06,Quay07,Yu05,Mizubayashi07}, but the results
remain inconclusive. Refs.~\cite{Utko06,Quay07} find no evidence for
electronic structures deviating from that of empty CNT quantum dots,
whereas Refs.~\cite{Yu05,Mizubayashi07} showed irregular
diamond-structures which were suggested to derive from the
encapsulated C$_{60}$ system. Since, at present, simultaneous
imaging and transport measurements are not possible, these
experiments may have probed peapods with rather different electronic
structure. For a consistent picture to emerge, more experiments on
high-quality peapod samples are clearly necessary. Thus the question
still remains whether the peapod system can provide new, interesting
and potentially useful functionality for nano-electronic circuitry?

We report a new set of detailed low-temperature transport
measurements for a peapod quantum dot device in the weakly coupled
Coulomb blockade regime. As in Refs.~\cite{Utko06,Quay07} we observe
well-defined Coulomb diamonds reflecting the discrete charging of
the CNT. In contrast to Refs.~\cite{Utko06,Quay07} and to similar
measurements on empty CNT quantum dots, we observe a systematic
repetition of avoided crossings in a gate-range corresponding to
some 400 consecutive charge-states. This systematic feature results
from a weak hybridization with a weakly gated localized orbital, as
detailed comparisons with master equation calculations show. We
tentatively propose these signatures to relate to a short chain of
C$_{60}$-molecules inside the CNT, residing close to one electrode.

\begin{figure*}[t]
\begin{center}
\includegraphics[width=2\columnwidth]{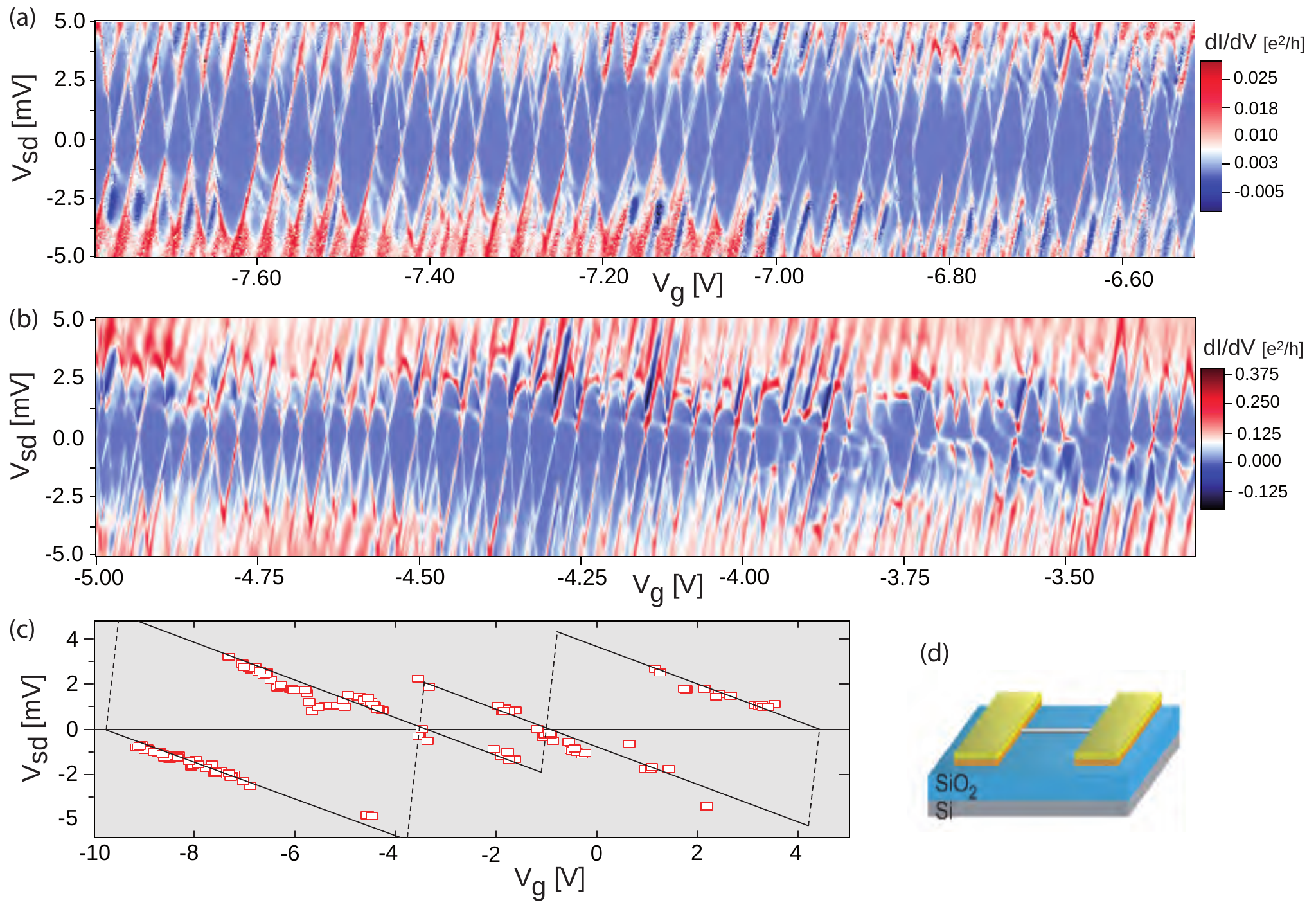}
\end{center}
\caption{
(a) and (b) Stability diagram, i.e., conductance ($dI/dV_{sd}$) as a
function of source-drain bias ($V_{sd}$) and gate-voltage ($V_g$)
at $300$~mK, showing a regular Coulomb blockade diamond pattern with
four-electron shell structure throughout the measured gate-range.
Diamonds are perturbed by a weakly gate-dependent feature
superimposed on the entire structure. (c) Observed avoided crossings
over the entire gate-range (red rectangles). Black lines are guides
to the eye, outlining the edges of the "impurity diamond". (d)
Sketch of the peapod quantum dot device.\label{fig:figure1}}
\end{figure*}

The single walled C$_{60}$ peapods of purity grade 90-95\%
(see~Ref.~\onlinecite{Utko06} for synthesis details), were suspended
in dichloroethane by sonification and dispersed in the form of
droplets onto an isolating SiO$_{2}$ layer of thickness 500\,nm,
thermally grown on top of a highly doped silicon substrate. By use
of atomic force microscopy imaging, individual tubes were identified
and then contacted by evaporated source and drain Ti/Au-electrodes
(25\,nm/25\,nm) using e-beam lithography. The device layout,
including the electrodes separated by $L\approx 600$~nm, is shown
schematically in Fig.~\ref{fig:figure1}d.

We have performed electronic transport measurements down to 300~mK
in a $^3$He cryostat, using standard lock-in techniques (AC source
drain voltage 50~$\mu$V RMS). Sweeping the gate-voltage $V_{g}$ and
measuring the linear conductance, we observe Coulomb blockade peaks
in metallic peapod samples~\cite{Utko06}. Here we concentrate on a
single sample exhibiting highly regular Coulomb blockade peaks in
the region -10~V$<V_{g}<$5~V, representative gate-ranges being shown
in Fig.~\ref{fig:figure1}a, b. We observe a clear four-electron
shell structure similar to that of empty
CNTs~\cite{Buitelaar02,Liang02,Sapmaz05}.

Unlike the device measured in Ref.~\onlinecite{Mizubayashi07}, which
also exhibited traces of a four-electron shell, there is no reason
to believe that the device studied here has been accidentally
partitioned into smaller sub-systems. In
Ref.~\onlinecite{Mizubayashi07}, the presence of distinct
gate-voltage regions with rather different, and surprisingly large
diamond sizes ($E_{add}\simeq 10-20$~meV for a peapod of length
$500$~nm), was interpreted as the tube being separated into two or
more smaller 'dots'.
\begin{figure}[t]
\begin{center}
\includegraphics[width=\columnwidth]{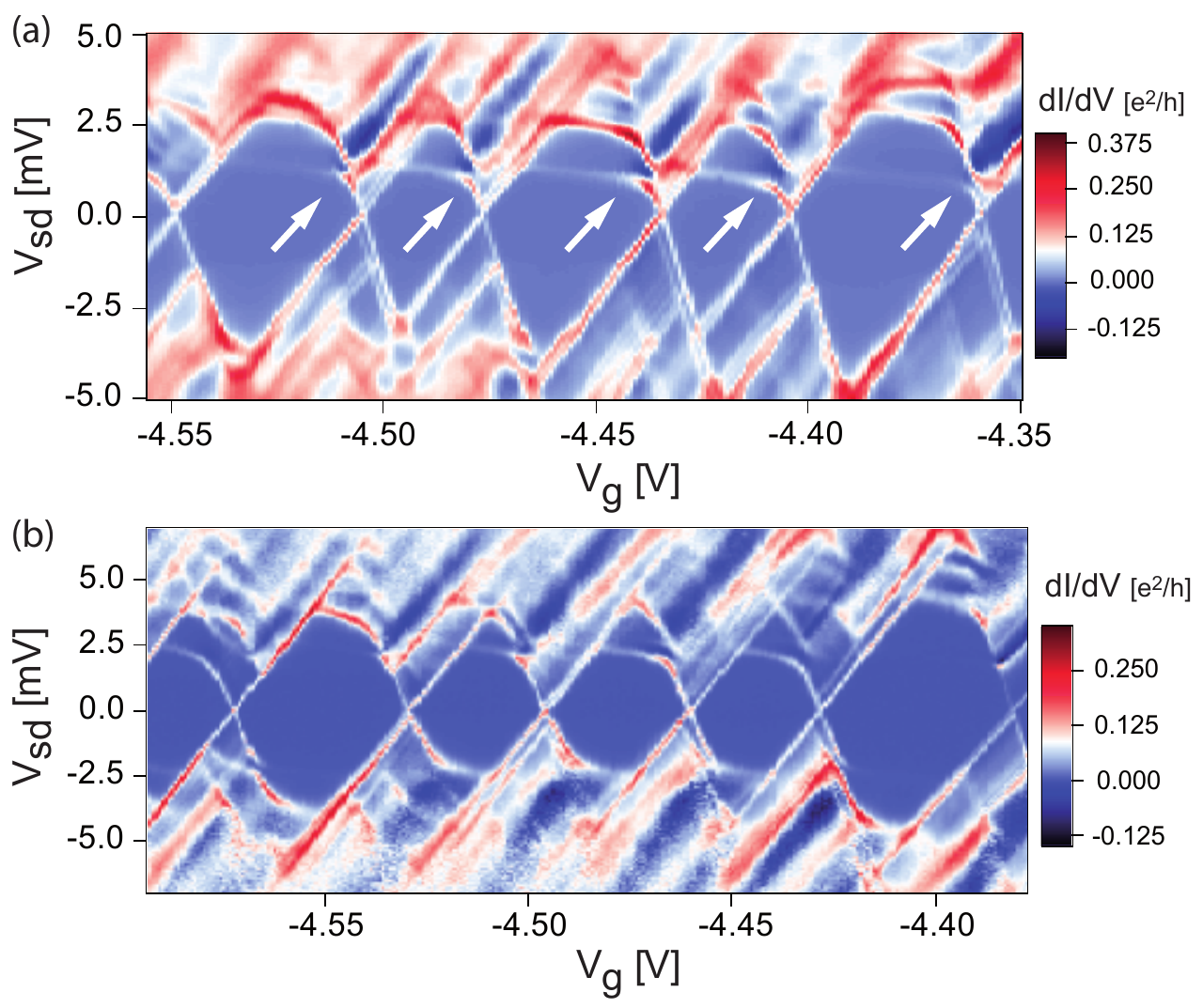}
\end{center}
\caption{ (a) Zoom in on a representative range of gate-voltages in
Fig.~\ref{fig:figure1}b. At positive bias we observe a series of
avoided crossings with a line of much lower gate-coupling than the
main diamond edges. (b) The same device after suspension. Avoided
crossings are seen at both negative and positive bias in the
displayed gate-range.\label{fig:figure2}}
\end{figure}

Having established the salient quantum dot features of these
transport data presented in Fig.~\ref{fig:figure1}a, b as essentially
CNT-like, one notices a distinct perturbation of the entire
stability-diagram: a very weakly gate-dependent resonance line
passes through the diamonds at positive and/or negative bias,
depending on the gate-voltage, and anti-crosses with the edges of
the CNT Coulomb diamonds. Fig.~\ref{fig:figure1}c shows the gate- and
bias-positions of these avoided crossings over the entire gate-range
where such features were observed. The weakly gate-dependent
resonance and the associated avoided crossings are seen more clearly in
Fig.~\ref{fig:figure2}a, which zooms in on a representative gate-range in
Fig.~\ref{fig:figure1}b. Due to the weak gate-dependence this line might
have been assigned to inelastic cotunneling~\cite{DeFranceschi01}.
This would, however, be inconsistent with the observed avoided
crossings with one side of the CNT diamond edges. Furthermore, the
weakly gate-dependent resonance does not occur symmetrically at the
same energy at positive and negative bias, as inelastic cotunneling
resonances do, and for most gate-voltages it is only present either
at positive or negative bias-voltage and strongly perturbs the
single-electron tunneling (SET) region on the corresponding
bias-side, showing broad regions of negative/positive differential
conductance (NDC/PDC).

As will be further substantiated below, all of these observations
are instead consistent with SET through a state which:
({\it i}) is much weaker coupled to the back-gate than are the
levels of the CNT; ({\it ii}) hybridizes with the levels of the
CNT; ({\it iii}) has a significant capacitive and tunnel coupling
only to the source lead.
We refer to this state as an "impurity orbital" to emphasize the
general nature of the transport effect in the following analysis.
After this we will argue that the impurity consists of a short chain
of $C_{60}$ molecules inside the tube.

\begin{figure}[h!]
\includegraphics[width=0.75\columnwidth]{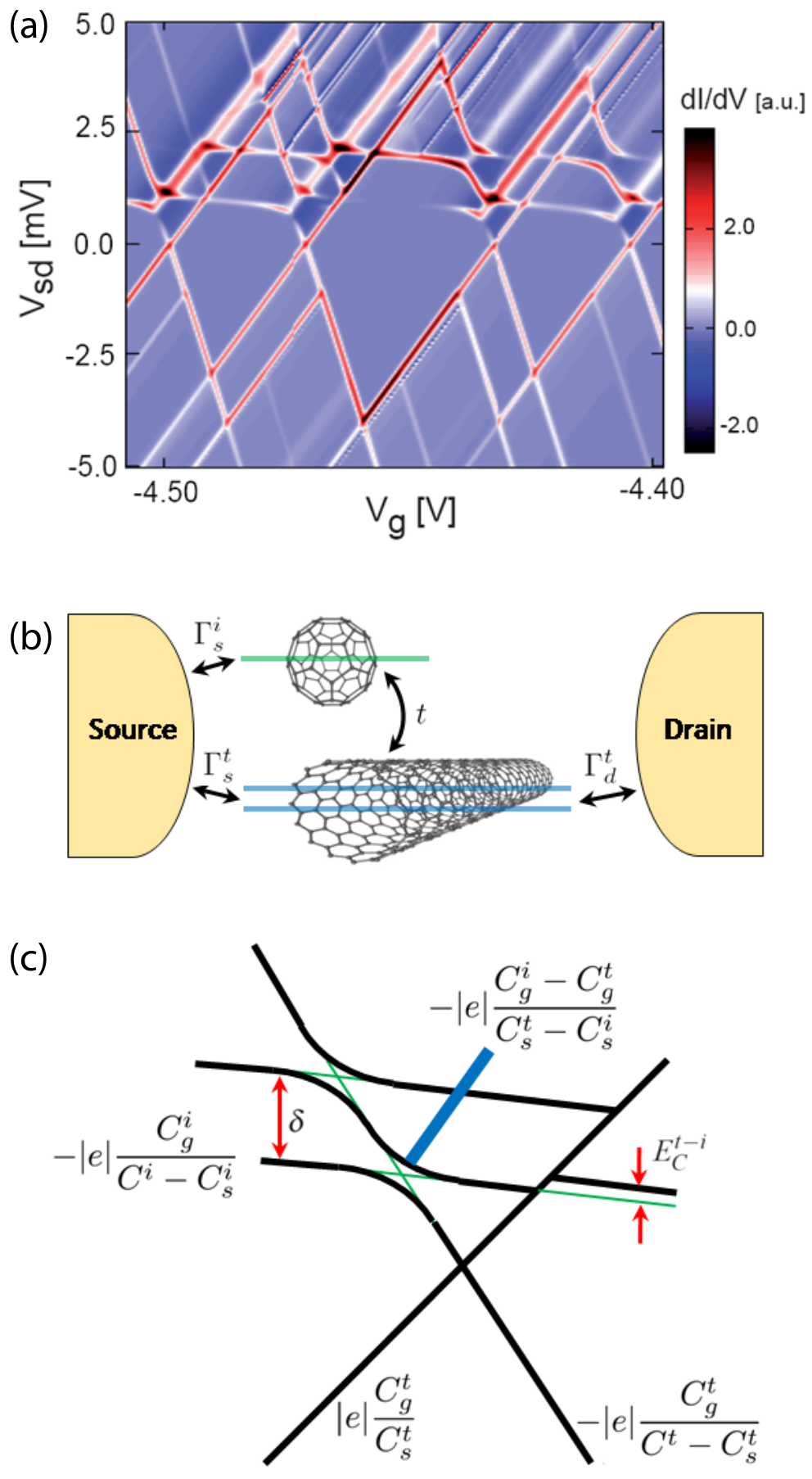}
\caption{\label{fig:figure3} (a) Calculated stability diagram
showing $dI/dV_{sd}$ as a function of source-drain bias ($V_{sd}$)
and gate-voltage ($V_g$) for a gate-voltage range corresponding to
the three central diamonds in Fig.~\ref{fig:figure2}a. In the
calculation we use the experimental temperature $T=300$~mK, but
neglect tunnel broadening. Therefore all resonances are somewhat
sharper than in the experiment. (b) Model system used in the
calculation. The impurity level hybridizes with both CNT subbands
(amplitude $t$), but is only tunnel coupled to the source (rate
$\Gamma_i^s$). Both CNT subbands are coupled with the same rate to
source and drain ($\Gamma_t^s = \Gamma_t^d$). (c) Sketch of avoided
crossings. The capacitances associated with the tube and impurity
can be read off from the slope of the resonance lines far from the
avoided crossing.}
\end{figure}

Independent of the precise origin of the impurity we can extract
detailed information about this state and its coupling to the CNT.
Fig.~\ref{fig:figure3}a  shows the result of model calculations,
reproducing the transport features in the central part of
Fig.~\ref{fig:figure2}a.
We model the CNT plus impurity state as sketched in
Fig.~\ref{fig:figure3}b, including only the lowest two subbands of the
CNT due to the large level-spacing $\Delta E \sim E_C$. The other
parameters of the constant interaction
model~\cite{Mayrhofer08,Sapmaz05} are extracted from the experiment:
$E_C = 2.9$~meV (charging energy), $\delta = 1.2$~meV (subband
splitting) and $dU = 0$~meV (excess Coulomb energy). The exchange
coupling, $J$, is difficult to extract from the data since it can
only be observed in transitions involving excited states. We
therefore take $J = 0$~meV, having verified that other values do
not qualitatively change the result.

The avoided crossings in the experimental data indicate that both CNT
orbitals (subbands 1 and 2) hybridize significantly with the
impurity orbital. Due to the different gate- and bias-couplings of
the CNT and impurity states (see below), the many-body peapod
model (CNT plus impurity plus hopping) has to be diagonalized
exactly at each gate- and bias-voltage point. Standard master equations
(lowest order perturbation theory in the tunnel coupling
to the leads) then suffice to explain all the features.

In the experimental setup the voltage is applied only to the source,
while the drain lead is grounded, $V_{sd} = V_{s} \propto -|e|
\mu_{s}$, where $-|e|$ is the electron charge and $\mu_s$ the
chemical potential of the source electrode. The voltage dependence
of the energy $\epsilon_m$ of orbital $m =1,2$ (CNT) and $m=i$ (impurity),
is then given by:
\begin{equation}\label{eq:voltage_dependence}
    \epsilon_m \propto -|e| \alpha_{g}^m V_{g} -|e| \alpha_{s}^m V_{sd},
\end{equation}
where $\alpha^{m}_k = C^{m}_k / C^m$, with $C^m = C^m_{s} + C^m_{d}
+ C^m_{g}$ being the sum of the capacitances to the source, drain
and gate electrodes. Fig.~\ref{fig:figure3}c shows a sketch of a
small part of a diamond with one avoided crossing and indicates how
to read off the gate- and bias-couplings far from the avoided
crossing, where there is little mixing between the CNT and impurity
states. For the CNT (regular pattern of Coulomb diamonds) we find
$\alpha_{g}^{1(2)} = 0.104$, $\alpha_{s}^{1(2)}=0.298$ (the same for
both subbands). Although the impurity resonance (weakly
gate-dependent line) crosses the zero-bias line, e.g., at around
$V_{g} = -3.75$~V ("impurity degeneracy point", see
Fig.~\ref{fig:figure1}b), only one side of the "impurity diamond"
(the drain-resonance) can be resolved, which is not enough to
determine both the gate- and bias-couplings. Note that the steep
dashed lines (source-resonance) drawn in Fig.~\ref{fig:figure1}c do
not allow for a reliable reading of the slope. However, the
capacitances also enter in the slope of the broad NDC and PDC
features inside the SET region, see blue line in
Fig.~\ref{fig:figure3}c, which in general are not parallel to the
CNT diamond edges. As discussed below, these indicate a resonance
between CNT and impurity orbitals. From these, we estimate
$\alpha_{g}^i \approx 0.0055$ and $\alpha_{s}^i \approx 0.99$. The
latter value indicates that the impurity level is "pinned" to the
chemical potential of the source ($C_{s} \gg C_{d},C_{g}$) and must
therefore be localized much closer to this lead. For simplicity we
assume the CNT to be symmetrically coupled to the source and drain,
$\Gamma_{s}^t = \Gamma_{d}^t$. (The general transport features are
found not to change qualitatively if an asymmetry is introduced.)
The impurity tunnel rates can not be read off directly, but through
an extensive theoretical survey we find that the best agreement with
the experiment is for a source coupling of the order of the CNT
tunnel coupling and a negligible drain coupling ($\Gamma_{s}^i = 4
\Gamma_{s,d}^t$,  $\Gamma_{d}^i = 0$ was used to obtain the result
shown in Fig.~\ref{fig:figure3}a). This is consistent with the above
finding based on the capacitances that the impurity level is
localized close to the source. The "top" of the impurity diamond is
not clearly visible in Fig.~\ref{fig:figure1}a, b. Nevertheless,
judging from the guiding lines drawn in Fig.~\ref{fig:figure1}c, we
can estimate the impurity charging energy from the height of the
smaller middle diamond to be $E_{C}^i \approx 2.5$ meV. The
"impurity diamonds" in Fig.~\ref{fig:figure1}c show a distinct
even-odd effect, indicating a large level-spacing, $\Delta E^i
\approx 2.5$ meV, and we only include one impurity orbital in the
model calculation. The hybridization between impurity and nanotube
can be read off from the magnitude of the avoided crossing to be
$t\approx0.15$ meV.

Altogether, the agreement between the experiment,
Fig.~\ref{fig:figure2}a, and model calculation,
Fig.~\ref{fig:figure3}a, is striking. We now discuss and explain
several qualitative features seen in both these plots.
The unusual almost horizontal resonance line, passing through the
Coulomb diamonds at positive bias, results in pronounced
avoided crossings with the source SET resonance only (marked with white
arrows in Fig.~\ref{fig:figure2}a).
At this point states with different numbers of electrons on the
CNT and impurity hybridize strongly and the corresponding
resonance lines avoid each other.
Within the SET regime are broad regions of PDC and NDC extending
from the avoided crossings.
These are, in fact, further signatures of the resonant hybridization
along the entire thick solid line sketched in Fig.~\ref{fig:figure3}c.
This mixing of impurity and tube states gives rise to interference
terms in the tunnel rates, which in turn affect the conductance.
Importantly, these conductance features do not correspond to the
usual condition of a resonance between a dot and a lead chemical
potential, but rather an internal resonance of the peapod system.
The width of this resonance is set by the hybridization $t$ and the
tunnel rate asymmetry, rather than temperature or tunnel-broadening.
For the gate-voltages we focus on here, none of the above features
are seen at negative bias since, due to the impurity bias-coupling,
the states which anti-cross here are too high in energy to
participate in transport. Instead, a regular pattern of sharp
conductance lines are seen, as expected for a pure CNT system.
The avoided crossings show an even--odd effect with alternating
magnitude of the gap, e.g., the one in the large central diamond is
less pronounced than those in the small neighboring diamonds.
The reason is that a filled CNT subband hybridizes more strongly
with an empty impurity state, as compared to the case where either
one has an open shell.
Furthermore, the impurity line does not pass straight through
consecutive diamonds, but instead makes a small upward jump (when
increasing $V_g$) between two diamonds. This implies a finite
capacitive coupling between CNT and impurity states and therefore a
"CNT--impurity charging energy", which we estimate as $E_C^{t-i}
\approx 0.1$~meV, see Fig.~\ref{fig:figure3}c.
In contrast to standard SET, the low-bias magnitude of the impurity
conductance lines depend sensitively on the voltages and in
particular they become very weak far from the avoided crossing.
This observation corroborates our earlier conclusion that the
impurity is located near the source, and far away from the drain,
which also implies that the impurity state is only tunnel coupled to
the source and the resonance lines seen are due to tunneling from
the drain. Only because of the hybridization with the CNT orbitals
there is an effective voltage-dependent tunnel coupling to both
leads, which becomes weaker further away from the avoided crossing.
There are also higher lying
resonances, as seen especially in the large central diamond
and indicated in Fig.~\ref{fig:figure3}c.
The bias-voltage separation to the lower impurity resonance is equal
to the subband splitting, $\delta = 1.2$ meV, and these higher lying
impurity resonances therefore correspond to tunneling
into the impurity, while at the same time the CNT is excited by
transferring one electron to the higher subband. This is possible
since the hybridization between the individual subbands and the
impurity induces an effective coupling between the subbands.

Having determined the properties of the impurity state and its
coupling to the CNT, we now return to the question of its nature.
One possibility is that we are dealing with an accidental impurity
residing outside the tube. In fact, this would be somewhat similar
to the alternative explanation given at the end of
Ref.~\onlinecite{Huettel09}, where there were no fullerenes present.
However, after the measurements presented above, the device was
covered with PMMA and the center of the tube was suspended
by electron beam lithography followed by
wet etching in a buffered solution of hydrofluoric acid,
creating an approximately 75 nm deep trench in the SiO$_2$ layer.
The result of low-temperature transport spectroscopy measurements after
suspension are shown in Fig.~\ref{fig:figure2}b and display features very
similar to before suspension. Most importantly, the weakly
gate-dependent impurity resonance is still present after suspension.
It is clearly seen at both negative and positive bias in the
gate-range shown in Fig.~\ref{fig:figure2}b, allowing an estimate of
roughly $\sim 5$ meV as an upper bound for its charging energy. Also
the magnitude of the hybridization as well as the impurity tunnel
couplings remain essentially unaltered by the etching process. This
makes the scenario of an impurity outside the CNT less likely.

Another possibility would be that the impurity is in fact a segment
of the CNT close to the source lead, separated from the rest of
the tube by a local defect. However, our extensive model
calculations clearly show that the transport data can only be
reproduced if the CNT orbitals have a significant coupling to both
source and drain leads, on the same order as the impurity--source
tunnel coupling.

Although no compelling evidence can be claimed, we argue that the
above observations suggest that the impurity is in fact a chain of
C$_{60}$ molecules inside the CNT, residing close to the source
electrode (a single C$_{60}$ would have a much larger charging energy).
The close proximity to the source
(and to some extent the surrounding nanotube) electrostatically
shields this fullerene chain, resulting in the low gate-coupling.
\par
As mentioned above, some previous studies~\cite{Hornbaker02, Lu03}
have suggested tube--impurity hybridizations several orders of
magnitude larger than the $\sim 0.15$ meV observed here. However,
different studies predict very different hybridization strengths,
which is also expected to depend sensitively on the type of peapod
being measured. In the present case, the detailed transport analysis
clearly shows that we are not dealing with a chain of fullerenes
extending throughout the tube, but rather a limited chain close to
one tube-end. Additionally, the CNT is operated in the Coulomb
blockade regime, where electron-electron interactions play a
dominant role in determining the mixing of the CNT and impurity
many-electron states.

In conclusion, we have measured low-temperature transport through a
single-walled carbon nanotube peapod quantum dot. Anomalous weakly
gate-dependent resonances, which show avoided crossings with the
standard CNT Coulomb diamonds, originate from an impurity which is
coupled both capacitively and by tunneling to its host nanotube.
Such coupled quantum dot systems with different electrostatic
properties can arise in various nanoscale transport junctions and
the detailed study presented here is of general use for interpreting
spectroscopic data. For instance the data in~\cite{Osorio07} show
evidence of a similar weakly gated "impurity state" giving rise to
NDC as it crosses the main diamond edge. Close inspection of the
data in Ref.~\onlinecite{Quay07} also reveal features which may be
interpreted in the same context. In the present case, several
observations point at the impurity in fact being a chain of C$_{60}$
molecules inside the nanotube, which is further supported by similar
measured data in another peapod device. The experimental observation
of electrically controlled mixing between the host nanotube and
fullerenes is important since it opens up the possibility of using
the latter's degrees of freedom for applications~\cite{Benjamin06},
for instance via the induced spin-exchange coupling.

This work was supported by the Danish Agency for Science, Technology
and Innovation (J.~P.), DFG-Forschergruppe 912 (S.~S.), the European
Union under the FP6 STREP program CANEL (A.~E., J.~N.), FP7 STREP
program SINGLE (J.~P., K.~F., M.~L.) and DFG SPP-1243 (M.~L.,
M.~W.). We thank Pawel Utko and Laure No\'{e} for experimental
contributions.

\end{document}